\documentclass[a4paper,11pt]{article}

\usepackage{jheppub} 

\usepackage[T1]{fontenc} 
\usepackage{mathabx}

\usepackage{multicol}

\newcommand\beq{\begin{equation}}
\newcommand\eeq{\end{equation}}
\newcommand\bal{\begin{aligned}}
\newcommand\eal{\end{aligned}}

\def\bF{\boldsymbol{F}}
\def\bA{\boldsymbol{A}}
\def\bB{\boldsymbol{B}}
\def\bC{\boldsymbol{C}}

\def\bJ{\boldsymbol{J}}
\def\bP{\boldsymbol{P}}
\def\bH{\boldsymbol{H}}
\def\bK{\boldsymbol{K}}
\def\bG{\boldsymbol{G}}
\def\bQ{\boldsymbol{Q}}

\def\bmF{\boldsymbol{\mathcal  F}}
\def\bmA{\boldsymbol{\mathcal A}}
\def\bmtA{\boldsymbol{\mathcal{\tilde{A}}}}

\def\bmJ{\boldsymbol{\mathcal J}}
\def\bmP{\boldsymbol{\mathcal P}}

\def\blambda{\boldsymbol{\lambda}}

\def\A{\hat A}
\def\B{\hat B}
\def\C{\hat C}
\def\D{\hat D}

\def\hmu{\hat\mu}
\def\hnu{\hat\nu}
\def\hrho{\hat\rho}

\def\halpha{\hat\alpha}

\def\hgamma{\hat\gamma}

\def\hsigma{\hat\sigma}

\def\n{n}

\title{\boldmath Fracton gauge fields from higher-dimensional gravity}

\author{Francisco Pe\~na-Ben\'itez\,}
\author{and\, Patricio Salgado-Rebolledo}
\affiliation{\vspace{.1cm}
Institute of Theoretical Physics,
Wroc\l{}aw University of Science and Technology,\\50-370 Wroc\l{}aw, Poland
\vspace{.3cm}}

\emailAdd{francisco.pena-benitez@pwr.edu.pl} \emailAdd{\kern1.45truecm 
patricio.salgado-rebolledo@pwr.edu.pl}

\abstract{We show that the fractonic dipole-conserving algebra can be obtained as an Aristotelian (and pseudo-Carrollian) contraction of the Poincar\'e algebra in one dimension higher. Such contraction allows to obtain fracton electrodynamics from a relativistic higher-dimensional theory upon dimensional reduction. The contraction procedure produces several scenarios including the some of the theories already discussed in the literature. 
A curved space generalization is given, which is gauge invariant when the Riemann tensor of the background geometry is harmonic.}

\begin{document} 
\maketitle
\flushbottom

\section{Introduction}

Fracton phases of matter represent a remarkable class of quantum states with novel and intriguing properties, challenging conventional paradigms in condensed matter physics \cite{Nandkishore:2018sel,Pretko:2020cko,Gromov:2022cxa,Grosvenor:2021hkn,Gorantla:2021svj}. These exotic phases are characterized by their restricted mobility of excitations, leading to unconventional patterns of long-range entanglement and topological order \cite{Vijay:2016phm,Bravyi:2011,Yoshida:2013sqa}. Understanding and describing the nature of fracton phases has emerged as a forefront of research, first in the fields of condensed matter theory and more recently in  high energy theory, promising groundbreaking insights into the behavior of quantum matter.

Gauge theories have been instrumental in describing a wide array of physical phenomena, from the fundamental forces of nature to condensed matter systems. Actually, to delve into the fascinating realm of the so-called gapless fracton phases, it is crucial to explore the role of symmetric gauge fields \cite{Pretko:2016lgv,pretko2019crystal,Pena-Benitez:2021ipo,Bidussi:2021nmp,Jain:2021ibh,Bertolini:2023sqa,Gorantla:2021svj}. These gauge fields mediate the interactions between fractonic matter.  
Therefore, the inclusion of symmetric gauge fields enriches the theoretical framework, paving the way for an insightful analysis of fracton phases in various materials and scenarios.
On the other hand,  fractonic theories  couple naturally to Aristotelian geometries \cite{Figueroa-OFarrill:2018ilb,deBoer:2020xlc,Pena-Benitez:2021ipo,Bidussi:2021nmp,Jain:2021ibh,glorioso2021breakdown,Armas:2023ouk}, i.e., manifolds whose tangent space isometry group is given only by rotations and spacetime translations, but no boosts as in the more familiar case of Riemann-Cartan geometry. Recently, generalizations of fracton electrodynamics defined on curved space have been constructed by gauging the Monopole Dipole Momentum Algebra (MDMA) \cite{Pena-Benitez:2021ipo,Bidussi:2021nmp,Jain:2021ibh,Armas:2023ouk}.

The main results of this paper are that the dipole conserving symmetry group can be embedded in Poincar\'e, and that symmetric gauge fields  can be obtained from a relativistic non-Einstenian gravity theory in one dimension higher. The higher-dimensional theory is described by a  Yang-Mills-like action with Poincar\'e as gauge group. Nonetheless, the theory is assumed to be in a Higgs phase with symmetry breaking pattern $\mathfrak{iso}(d+1,1)\to \mathfrak{so}(d,1)$.
The fracton gauge fields are obtained after a particular limit defined by a Lie algebra contraction that connects the Poincar\'e algebra with the MDMA. Such contraction is a suitable combination of  pseudo-Carrollian and Aristotelian limits respectively. Additionally by dimensionally reducing the system we identify the extra dimension of the system with the internal $U(1)$ associated to monopole transformations. Similarly, the dipole transformations in the dimensionally reduced theory is obtained from the spacetime boosts along the compactified spatial dimension.

The structure of this paper is as follows: In Section \ref{sec:dipolecon}, we provide a brief overview of dipole conserving systems, highlighting their defining characteristics.  In addition, we introduce symmetric gauge theories as the fields mediating the interactions between fractons, and discuss the incompatibility between the gauge principle and spacetime curvature.
In Section \ref{sec:TheTheory} we define a novel contraction of the Poincar\'e algebra that leads to the MDMA algebra in one dimension less. Then, we define a higher-dimesional Riemann-squared gravitational theory, and after the algebra contraction we dimensionally reduce it. After doing so, we discuss the results and finish with some conclusions and  comments on possible future developments  in Section \ref{sec:conclusions}.

\section{Symmetric gauge fields and dipole conservation}
\label{sec:dipolecon}
Before discussing  dipole conserving systems, let us consider a spinless particle in $d+1$ spatial dimensions with momentum $P_A$, and angular momentum $J_{AB}=x^AP_B - x^B P_A$. If the system is translational and rotational invariant $P_A$ and $J_{AB}$ will be constants of motion, which implies that
\begin{align}
     \dot P_A &= 0,\\
     \dot x^A & \propto P_A.
\end{align}
Actually notice that the conservation of angular momentum does not introduce new constants of motion, instead it constraints the particles's motion by requiring the velocity to be parallel to the momentum.  Now let us pick coordinates  $x^A=(x^i,R\, z)$, interpret $x^i$ as the coordinates of the physical space, $z$ the coordinate of some extra dimension, and $R$ a constant with dimensions of length. Then, we  redefine the transverse momentum as $Q \equiv R P_z$, and the transverse angular momentum $Q^i \equiv R J_{i\,z}$. After doing so, we zoom out the $z-$direction  by sending $R\to 0$. Therefore, the angular momenta become
\begin{align}
    J_{ij} &=x^iP_j - x^j P_i,\\
    Q^i &=Qx^i.
\end{align}
From the lower dimensional perspective the transverse momentum takes the form of an internal charge, and the angular momentum $Q^i$ corresponds to the dipole moment of that charge $Q$. For such a particle the conservation of all the charges imply
\begin{align}
     P_i &= \mathrm{constant},\\
     x^i &= \mathrm{constant},\\
     J_{ij} &= 0,
\end{align}
which is unusual since the particle is not allowed to move, however, its momentum is not constrained to be related to the velocity. 
Although a system with such properties seems to be dynamically trivial, we notice that motion is allowed once more than one particle are included. For example, let us consider two fractons with charges $Q_1=q$, $Q_2=-q$ such that $Q=0$, and $Q^i=q( x_{(1)}^i-x_{(2)}^i)$ for such a system the conservation of momentum and dipole imply
\begin{align}
     P_{(1)i}  & = \frac{1}{2}\left( P_i + W_i(t)\right), \qquad x_{(1)}^i   =  X^i(t) + \frac{1}{2q}Q^i  \,\\
     P_{(2)i}  & = \frac{1}{2}\left(P_i - W_i(t)\right), \qquad x_{(2)}^i   =  X^i(t) - \frac{1}{2q}Q^i  .
\end{align}
Thus the actual dynamical variables for a dipole are the center of mass position $ X^i(t)$, and the relative momentum $W_i(t)$ between the particles forming the dipole. If in addition we impose conservation of angular momentum we obtain the condition
\begin{equation}
    \dot X^{[j}(t)P^{k]} =  \frac{1}{2q}\dot W^{[j}(t) Q^{k]},
\end{equation}
which for $d=3$ constraints the force $\dot W^i$ as
\begin{align}
    \dot X^i(t) & \equiv V^i(t),\\
    \dot W_i(t) & = \gamma(t)Q^i  - \frac{4q^2}{|\mathbf Q|^2}Q^j( V^i(t)P_j -V^j(t)P_i)\,.
\end{align}

In fact, we can go beyond the single particle picture assuming locality and introducing the densities $\varrho,p_i$ such that we can express the conserved charges as 
\begin{align}
    Q &=\int d^d x\, \varrho, \qquad  Q^i= \int d^d x\, x^i \varrho, \label{eq:QandQi}\\
    P_i &=\int d^d x\, p_i, \qquad J_{ij}= \int d^d x\, \left( x^i\,p_j - x^j\,p_i\right) .
\end{align}
With these definitions and requiring  $\dot Q = 0$,  $ \dot P_i = 0 $, we obtain the continuity equations
\begin{align}\label{eq_conser1}
    \partial_0\varrho+\partial_i j_i &=0,\\
    \partial_0 p_j+\partial_i \tau_{ij} &=0.
\end{align}
On the other hand, dipole and angular momentum conservation $\dot Q^i =0$, and  $\dot J_{ij} =0$ demands the constraints
\begin{align}
    j_i & = \partial_j K_{ji},\\
    \tau_{[ij]} & = \partial_k L_{kij},
\end{align}
with $L_{ijk}=-L_{ikj}$, and $K_{ij}$ the dipole current. However, without lost of generality the stress tensor can be improved such that the new stress tensor $T_{ij}$ is symmetric. With such improvement the conservation of angular momentum is automatic if momentum is conserved. Similarly, to guarantee conservation of both charge and dipole moment it is enough to satisfy the generalized continuity equation\footnote{Notice that the antisymmetric part of $K_{ij}$ drops from Eq. \eqref{eq_conser2}}
\begin{align}\label{eq_conser2}
    \partial_0\varrho+\partial_i\partial_j K_{ij} &=0.
\end{align}
\subsection{Fracton Gauge Theory}

In fact, the charge (dipole) conservation can be derived once the scalar $\phi$, and symmetric tensor $A_{ij}$ gauge fields are   minimally coupled to the fracton current as
\begin{equation}
    S\sim \int d^{d+1}x \left(\varrho\phi + K_{ij}A_{ij}\right),
\end{equation}
with the gauge field transforming as
\beq\label{gtflat}
\delta A_{ij} = \partial_i \partial_j \varepsilon,\quad
\delta \phi = -\dot\varepsilon.
\eeq
Actually, in a system with dynamical gauge fields, the corresponding  generalized electrodynamic theory with electric and magnetic fields defined as \cite{Pretko:2016lgv}

\beq\label{curvaturesflat}
F_{ijk}=\partial_i A_{jk} -\partial_j A_{ik},\quad
F_{0ij}=\dot A_{ij}+\partial_i\partial_j \phi,
\eeq
with action 
\beq \label{ActionPretko}
S=\frac{1}{2}\int d^{d+1}x\left( F_{0ij} F_{0ij} -\frac12 F_{ijk} F_{ijk} \right).
\eeq

Notice that it is possible to construct a symmetric spacetime tensor $A_{\mu\nu}$ with $A_{0\mu} = -\partial_\mu\phi$, and  transformation law $\delta A_{\mu\nu} = \partial_\mu\partial_\nu\varepsilon$. The field strength for the enhanced model is defined as
\begin{equation}
F_{\mu\nu\rho}=\partial_\mu A_{\nu\rho} -\partial_\nu A_{\mu\rho},    
\end{equation}
and the action \eqref{ActionPretko} can be written as
\beq\label{ActionPretkoCov}
S=\frac{1}{4} \int d^{d+1}x F_{\mu\nu\rho} F^{\mu\nu\rho},
\eeq
where indices are raised and lowered with the Minkowski metric $\eta_{\mu\nu}={\rm diag}(-+\cdots+)$. Notice, that such construction posses and accidental gauge symmetry $\delta A_{\mu\nu}=\partial_\mu\beta_\nu$ that leaves the field strength invariant but does not respect the symmetry property of the gauge field. Actually we notice that the alternative gauge field $\tilde A_{\mu\nu} = A_{\mu\nu} + \partial_\mu\phi\,\tau_\nu$ with $\tau_\nu=\delta^0_\nu$ has field strength $\tilde F_{\mu\nu\lambda} = F_{\mu\nu\lambda}$, therefore, same equation of motions. It is important to  emphasize that the constraint $\tau^\mu A_{\mu\nu}=\partial_\mu\phi$ explicitly breaks the apparent relativistic symmetry.

\subsection{Fractons on curved manifolds}
\label{sec:action}

In order to illustrate the tension between the fractonic gauge principle and curved spaces, let us consider a Minkowskian manifold  with  metric $g_{\mu\nu}$ and a timelike vector $\tau^\mu$. We can define a spatial metric by introducing a clock $\tau_\mu$ as
\beq
h_{\mu\nu}= \tau_\mu \tau_\nu + g_{\mu\nu},\qquad \tau_\mu \tau^\mu=1, \quad h_{\mu\nu}\tau^\nu=0,
\eeq
In addition, we introduce an Aristotelian covariant derivative  with the  torsion-free connection
\beq\label{Arconnection}
\nabla_\mu \tau_\nu=0=\nabla_\mu h_{\nu\rho}\,\qquad
\Gamma^\mu_{\;\;\nu\rho}= \tau^\mu \partial_{\nu} \tau_{\rho}+ \frac12 h^{\mu\lambda}\left(
\partial_\nu h_{\lambda \rho}
+\partial_\rho h_{\nu\lambda}
-\partial_\lambda h_{\nu \rho}
\right),
\eeq
and $\partial_\mu\tau_\nu-\partial_\nu\tau_\mu=0$.
With these ingredients at hand, we consider the symmetric tensor gauge field $A_{\mu\nu}$ of the previous section   satisfying the covariant condition
\beq
\label{constraintflat}
\tau^\nu A_{\mu \nu}= -\partial_\mu \phi,
\eeq
under gauge transformations we postulate the field $A_{\mu\nu}$ transforms as\footnote{Notice that a covariant theory related to fractons for a symmetric rank-two gauge field with the same type of gauge transformation has been considered in \cite{Bertolini:2022ijb,Bertolini:2023sqa}. However the form of the field strength and therefore the action are different.}
\beq\label{curvedGT}
\delta A_{\mu\nu} =  \nabla_\mu  \nabla_\nu \epsilon,
\eeq
and  its corresponding field strength reads
\beq
F_{\mu\nu\lambda}=\nabla_\mu A_{\nu\lambda} -\nabla_\nu A_{\mu\lambda}.
\eeq
The issue this construction has is that the field strength is not gauge invariant, in fact it can be shown that
\beq
\delta F_{\mu\nu\rho} = [\nabla_\mu,\nabla_\nu]\partial_\rho\epsilon = -R^\alpha\,_{\rho\mu\nu}\,\partial_\alpha\epsilon.
\eeq
Therefore, a minimal extension of the action \eqref{ActionPretkoCov}  can be
\beq\label{Actioncurvedimp}
S=-\int d^{d+1}x\;\sqrt{|g|}\;
\left(\frac12 F_{\mu\nu\lambda} F^{\mu\nu\lambda}
- R^{\mu\nu\rho\sigma} A_{\mu\rho} A_{\nu\sigma} 
\right).
\eeq
However, this action is not invariant in arbitrary backgrounds. In fact, it  preserves gauge invariance  for background spacetime with curvature satisfying
\beq
\nabla_\mu R^{\mu\nu\rho\sigma}=0.
\eeq

In addition, notice that the theory can be reformulated using frame fields $e^a_{\;\,\mu}$ such that $h_{\mu\nu}=\delta_{ab}\, e^a_{\;\,\mu}e^b_{\;\,\nu}$, and  satisfying $\nabla_\mu e^a_{\;\,\nu} + \omega^{ab}_{\;\,\;\,\mu} e_{b\nu}=0$ with $\omega^{ab}_{\;\,\;\,\mu} $ the $\mathfrak{so}(d)$ connection. Moreover, we introduce the inverse spatial vielbein $e_a^{\;\,\mu}$ and define $A^a_{\;\,\mu}=A_{\mu\nu} e^{a\nu}$. Therefore the gauge field $A_{\mu\nu}$ can be split as
\beq\label{Amunu}
A_{\mu\nu}= A^a_{\;\,\mu} e_{a\nu}-\partial_\mu \phi \,\tau_\nu
\eeq
where we have used the constraint \eqref{constraintflat}. Besides, the field $A^a_{\;\,\mu}$ satisfies
\beq\label{projA}
A_{a\mu} \tau^\mu =-\partial_\mu \phi \,e_a^{\;\,\mu},\qquad
A_{a\mu} e_b^{\;\,\mu}= A_{\mu\nu} e_a^{\;\,\mu} e_b^{\;\,\nu}\equiv A_{ab},
\eeq
and thus can be put in the form
\beq\label{Aamu}
A_{a\mu} = -\partial_\nu \phi \,e_a^{\;\,\nu} \tau_\mu + A_{ab} e^b_{\;\,\mu}.
\eeq
Using the fact that the Riemann tensor constructed out of the Aristotelian connection \eqref{Arconnection} satisfies $R^{\mu\nu}_{\;\;\;\;\rho\sigma}\tau_\mu=0$, and the field strength of $A_{\mu\nu}$ satisfies $F_{\mu\nu\rho} \tau^\rho=0$ the action \eqref{Actioncurvedimp} can be written in terms of $A_{a\mu}$  as 
\beq\label{Actioncurvedimp1}
S= -  \int d^{d+1}x\, \sqrt{|g|}\, 
\bigg[ \frac12 \tilde F^a_{\;\,\mu\nu}
\tilde  F_{a}^{\;\,\mu\nu}
-
R^{ab\mu\nu} A_{a\mu} A_{b\nu}
\bigg].
\eeq
where 
\beq
\tilde F^a_{\;\,\mu\nu} \equiv F_{\mu\nu\rho}e^{a\rho}= D_{[\mu} A^a_{\;\nu]},
\eeq
and $D_\mu$ is the (rotational) covariant derivative built up with $\omega^{ab}_{\;\,\;\,\mu}$. The transformation law of $A^a_{\;\,\mu}$ that is compatible with \eqref{curvedGT} reads
\beq\label{transformationAamu}
\delta A^a_{\;\,\mu}=D_\mu \left(e^{a\nu}\partial_\nu\epsilon\right).
\eeq

Alternatively to the action \eqref{Actioncurvedimp},  it is possible to introduce a Higgs mechanism for the dipole symmetry as shown in Ref. \cite{Pena-Benitez:2021ipo}, by doing so a gauge invariant action can be obtained after introducing a St\"uckelberg field $\psi_\alpha$ transforming as $\delta\psi_\alpha=\partial_\alpha \epsilon $ such that the new gauge field is gauge invariant
\beq
\bar A_{\mu\nu} = A_{\mu\nu}-\nabla_\mu \psi_\nu
\eeq
and therefore its field strength
\beq
\bar F_{\mu\nu\rho} = \nabla_\mu \bar A_{\nu\rho}
-\nabla_\nu \bar A_{\mu\rho}
=F_{\mu\nu\rho} + R^\alpha_{\;\;\rho\mu\nu}\,\psi_\alpha\,
\eeq
is gauge invariant, and no constraints  on the background spacetime are needed. Therefore, we define the action
\beq
S=-\frac12 \int d^{d+1}x\;\sqrt{-g}\;
\bar F_{\mu\nu\lambda} \bar F^{\mu\nu\lambda}.
\eeq
Since $R^\alpha_{\;\;\mu\nu\rho}\tau_\alpha=0$, the longitudinal component of the St\"uckelberg field $\psi_\mu\tau^\mu$ does not appear in the action, therefore, it will be completely undetermined. In fact, if we fix  $\psi_\mu$ to be of the form
\beq
\psi_\mu = -\phi \tau_\mu +\psi^a e_{a\mu},
\eeq
and use Eq. \eqref{Amunu} we can write the invariant gauge field as
\beq
\bar A_{\mu\nu}= 
\left(A^a_{\;\,\mu}-D_\mu \psi^a\right)e_{a\nu}.
\eeq
With that choice for the St\"uckelberg the action reduces to 
\beq\label{Actioncurvedimp2}
S
=-\frac12 \int d^{d+1}x\;\sqrt{-g}\;
\left( \tilde F^a_{\;\,\mu\nu}-R^{ab}_{\;\,\;\,\mu\nu}\psi_b \right) \left(\tilde F_a^{\;\,\mu\nu}-R_{ac}^{\;\,\;\,\mu\nu}\psi^c \right),
\eeq
which corresponds to the action obtained in \cite{Pena-Benitez:2021ipo} after gauging the dipole conserving symmetry group.

In the next section, we will define the 
Monopole Dipole Momentum Algebra and embed it into the Poincare group in one extra dimension and
show that the actions \eqref{Actioncurvedimp}, \eqref{Actioncurvedimp1}, \eqref{Actioncurvedimp2} can be obtained from a gravitational gauge theory upon dimensional reduction after applying an Aristotelian limit.

\section{From Poincar\'e gauge theory to fracton gauge fields}\label{sec:TheTheory}

Following the intuition built in Sect. \ref{sec:dipolecon} we will embed the fracton algebra into the Poincar\'e algebra $\mathfrak{iso}(d+1,1)$ with the purpose of deriving the fractonic gauge theory as some limit of a gauge theory of gravity in one dimension higher.

\subsection{Fractonic Symmetry Algebra}

In systems with dipole conservation the charge conservation cannot be described as a regular internal $U(1)$, since the total value of the dipole moment in the system changes once a space translation is applied (see Eq. \eqref{eq:QandQi}).  Actually, this property is consequence of the generators of space translations and the transformation generated by the dipole charge not commuting, and forming a non-Abelian symmetry  algebra \cite{Gromov:2018nbv,Grosvenor:2021rrt}
\beq\label{Halg}
[ \bP_i, \bQ^j]= \delta_i^j  \bQ,
\eeq
where $\bP_i$, $\bQ^j$, $\bQ$ are the generators of translations, "dipole", and $U(1)$ transformations. In addition,
if the system possesses
rotational invariance, the bracket Eq. \eqref{Halg} has to be supplemented with
\beq\label{JPQ}
[  \bJ_{ij},  \bQ^k] = \delta^k_{[j} \bQ_{i]},\qquad
[  \bJ_{ij},  \bP_k ]= \delta_{k[j} \bP_{i]},\qquad
[  \bJ_{ij},  \bJ_{kl} ]= \delta_{[i[k}\bJ_{l]j]}, 
\eeq
where $\bJ_{ij}$ is the generator of the $\mathfrak{so}(d)$ rotation group.  We will refer to this algebra as Monopole Dipole Momentum Algebra (MDMA).

The commutation relation Eq. \eqref{Halg} has important implications on the properties of fractons systems on curved space. In fact, as we discussed in the previous section spacetime curvature is in tension with the fractonic gauge transformations, and in particular the origin of the symmetry breaking has been argue to have the same origin as the translational invariance breaking in gauge theories of spacetime symmetry groups \cite{Pena-Benitez:2021ipo}. 

We also notice that MDMA is isomorphic to    the Carroll algebra \cite{Bacry:1968zf,Marsot:2022imf} 
\beq\label{Carroll}
\bal
&[ \bP_i, \bC^j]= \delta_i^j \bH,&\qquad&
[  \bJ_{ij},  \bC^k] = \delta^k_{[j} \bC_{i]},
\\&
[  \bJ_{ij},  \bP_k ]= \delta_{k[j} \bP_{i]},&\qquad&
[  \bJ_{ij},  \bJ_{kl} ]= \delta_{[i[k}\bJ_{l]j]},
\eal
\eeq
which requires to reinterpret the $U(1)$ generator $\bQ$ as the Hamiltonian $\bH$ and the dipole generator $\bQ^i$ and the generator of Carrollian boosts $\bC^i$. The analogy between the dipole conserving group and the Carroll symmetry has been extensively used to construct fractonic models of particles and gauge fields \cite{Casalbuoni:2021fel,Marsot:2022imf,Figueroa-OFarrill:2023vbj}. A remarkable property of Carroll theories is that it corresponds to the vanishing speed of light limit of the Poincar\'e algebra \cite{Bacry:1968zf},  and are also characterized  by immobile excitations
\cite{Bergshoeff:2014jla,Marsot:2022imf,Figueroa-OFarrill:2023vbj}

The Carroll algebra can be obtained from the Poincar\'e algebra $[\bJ_{\mu\nu},\bP_\rho]=\eta_{\rho[\nu}\bP_{\mu]}$, $[\bJ_{\mu\nu},\bJ_{\rho\sigma}]=\eta_{[\mu[\rho}\bJ_{\sigma]\nu]}$, where $\mu=(0,i)$ and $\eta_{\mu\nu}={\rm diag}(-,+,\cdots,+)$, by means of the following Lie algebra contraction \cite{Bergshoeff:2017btm}
\beq\label{Ccontraction}
\mathcal \bP_0=\frac{1}{c}\bH,\qquad \mathcal \bJ_{0i}=\frac{1}{c} \bC_i,
\eeq
when the speed of light $c$ vanishes. This suggests that the dipole conserving algebra can be obtained by means of a similar limiting procedure from Poincar\'e. In the next section we show that this is indeed the case.

\subsection{Poincar\'e and the fracton symmetry algebra}
\label{sec:contraction}

In this section we will show how the MDMA in $d+1$ dimensions can be obtained after contracting Poincar\'e algebra in $\mathfrak{iso}(d+1,1)$ in $d+2$ dimensions, given by
\begin{equation}\label{Poincare}
[\bmJ_{ \A \B},   \bmP_{\C}] =  \eta_{\C [\B}  \bmP_{\A]}
,\hskip 2.5 truecm
[ \bmJ_{\A \B}, \bmJ_{\C \D}] = \eta_{[\A[\C} \bmJ_{\D] \B]},
\end{equation}
where $\eta_{\A \B}={\rm diag}(-+\cdots+)$ is the Minkowski metric and $\A=0,1,\dots, d,d+1\equiv \n$. The resemblance between the dipole conserving algebra and the Carroll algebra suggests that a rescaling analog to \eqref{Ccontraction} will allow to obtain the later as a Lie algebra contraction of Poincar\'e. However, temporal translations are also present in fractonic systems, where the generator $\bH$ commutes with all the generators of the dipole conserving algebra. Thus, instead of using time as the longitudinal direction in the contraction, we can use a spatial direction $\A=n$ to define a pseudo-Carrollian contraction\footnote{Similarly, a pseudo-Galilean contraction using a spatial direction instead of time as longitudinal direction has been considered in \cite{Hartong:2017bwq}.} similar to \eqref{Ccontraction}, i.e.
\beq\label{pseudoC}
\bmP_n=\frac1\sigma \bQ\,\qquad \bmJ_{An}=\frac1\sigma \bQ_A,\qquad A=(0,a) ,\qquad\sigma\rightarrow0.
\eeq
The extra dimension will be interpreted as the internal direction that will be associated to the conservation of monopole charge. However, the resulting symmetry will still be relativistic, which contrast with the absence of boost symmetry in the dipole conserving algebra and the Aristotelian character of fractonic systems. Thus, one should supplement \eqref{pseudoC} with a contraction of Aristotelian nature that eliminates transformations that connect space and time translations. This can be achieved through the rescaling
\beq\label{Aris}
\bmJ_{0A'}=\frac1\varepsilon \bG_{A'} ,\qquad A'=(a,n),\qquad \varepsilon\rightarrow0.
\eeq
Therefore,
we select the time and spatial direction $(0,n)$ and split the indices as $\A=(0,a,\n)$, where $a=1,\dots,d$. The commutation relations \eqref{Poincare} then take the form
\begin{multicols}{2}
\begin{subequations} \label{splitpoincare}
\setlength{\abovedisplayskip}{-14pt}
\allowdisplaybreaks
\begin{align}
&[\bmJ_{0\n}, \bmP_{\n}]= \bmP_{0},
\\
&[\bmJ_{a\n}, \bmP_{\n}]= \bmP_{a},
\\
&[\bmJ_{0\n}, \bmP_{0}]=\bmP_{\n},
\\
&[\bmJ_{a\n}, \bmP_{b}]=-\delta_{ab}\bmP_{\n},
\\
&[\bmJ_{0\n},\bmJ_{a\n}]= - \bmJ_{0a},
\\
&[\bmJ_{a\n},\bmJ_{b\n}]= - \bmJ_{ab},
\\
&[ \bmJ_{0a}, \bmJ_{0b}]= \bmJ_{ab},
\\
&[ \bmJ_{ab}, \bmJ_{0c}]=\delta_{c[b} \bmJ_{0a]},
\\
&[ \bmJ_{ab}, \bmJ_{cd}]=\delta_{[a[c} \bmJ_{d]b]},
\\
&[ \bmJ_{0a},\bmJ_{0\n}]= \bmJ_{an},
\\
&[ \bmJ_{0a},\bmJ_{b\n}]= \delta_{ab}\bmJ_{0\n},
\\
&[ \bmJ_{ab},\bmJ_{c\n}]= \delta_{c[b}\bmJ_{a]\n},
\\
&[ \bmJ_{0a}, \bmP_{0}]= \bmP_{a},
\\
&[ \bmJ_{0a}, \bmP_{b}]= \delta_{ab} \bmP_{0},
\\
&[ \bmJ_{ab}, \bmP_{c}]= \delta_{c[b} \bmP_{a]}.\\
\nonumber
\end{align}
\end{subequations}
\end{multicols}\noindent
Combining the Aristotelian contraction \eqref{pseudoC} and the pseudo-Carrollian contraction \eqref{Aris} lead us to define the following rescaling of the elements of the higher-dimensional Poincar\'e algebra
\beq\label{genresc}
\bal
&\bmP_0= \bH\, ,\quad
\bmP_a=\bP_a\, ,\quad
\bmP_{\n} =\frac{1}{ \sigma} \bQ\, ,\quad
\bmJ_{a\n} = \frac{1}{ \sigma}  \bQ_a, \quad
\bmJ_{ab} = \bJ_{ab},\\
&\bmJ_{0\n} = \frac{1}{ \varepsilon \sigma}    \bK,\quad \bmJ_{0a} = \frac{1}{ \varepsilon}  \bG_a,\quad
\quad
\eal
\eeq
where the generators in the first line of Eqs. \eqref{genresc} are to be interpreted as spacetime translations, $U(1)$, dipole transformations, and rotations respectively. Taking the limit $\varepsilon\rightarrow 0$, we find that the set of generators $\{\bJ_{ab},\bH,\bP_a,\bQ,\bQ_a\}$ close in a sub-algebra defined by the following non-vanishing commutators
\begin{multicols}{2}
\begin{subequations} \label{dipolealgext}
\setlength{\abovedisplayskip}{-14pt}
\allowdisplaybreaks
\begin{align}
&[\bP_a, \bQ_{b}]=\delta_{ab}\bQ,
\\
&[ \bJ_{ab}, \bP_{c}]= \delta_{c[b} \bP_{a]},
\\
&[ \bJ_{ab},\bQ_c]= \delta_{c[b}\bQ_{a]},
\\
&[ \bJ_{ab}, \bJ_{cd}]=\delta_{[a[c} \bJ_{d]b]},
\\
&[\bQ_a, \bQ]= \sigma^2 \bP_{a},
\\
&[\bQ_a,\bQ_b]= -\sigma^2 \bJ_{ab},
\end{align}
\end{subequations}
\end{multicols}\noindent
whereas the generators $\{\bG_a,\bK\}$ form an ideal with commutation relations
\beq\label{Ideal}
[ \bJ_{ab}, \bG_c]=\delta_{c[b} \bG_{a]},
\qquad[ \bG_a,\bQ_b]= \delta_{ab}\bK,
\qquad[\bQ_a,\bK]= \sigma^2\bG_a.
\eeq
The sub-algebra \eqref{dipolealgext} is an extension of the MDMA defined by \eqref{Halg} and \eqref{JPQ}, and reduces to it in the limit $ \sigma \rightarrow 0$. Contrary to the standard contractions where the parameters are removed from the algebra, we will take the strict limit $\varepsilon\to 0$, but keep the leading terms in $\sigma$ for reasons that will become clearer below.

\subsection{Poincar\'e Gauge Theory}

After showing that the  MDMA can be obtained after contracting the Poincar\'e algebra, we will proceed constructing a relativistic gravity theory in five-dimensions with Poincar\'e as gauge group, and then dimensionally reduce it. This analysis has the purpose of understanding the puzzling relation between fracton phases of matter and gravity theories.
As we will show, starting from a fully boost invariant action in higher dimensions is not enough to recover an invariant  fracton gauge theory. This is related to the fact that in the strict $\sigma\to 0$ limit the lower dimensional theory is purely gravitational without fracton gauge fields. Therefore, the gauge group will be realized non-linearly by adding a Higgs field $\Psi^A$ associated to the transverse boosts $\bmJ_{A n}$, where un-hatted capital indices take values $A =0,\dots,d$. Therefore, the  non-linear connection one-form $\bmA =\bmA_{\hmu} dx^{\hmu}$, $\hmu=0,\dots,d+1$, taking values on the $\mathfrak{iso}(d+1,1) $ algebra Eq. \eqref{Poincare} reads
\beq\label{relconnection}
\bmA=  E^{\A} \bmP_{\A} +\frac{1}{2} \Omega^{\A \B} \bmJ_{\A \B},
\eeq
where $E^{\A}_{\;\;\mu}$ and $\Omega^{\A \B}_{\;\;\;\;\mu}$ are the non-linear vielbein and spin-connection, respectively.  On the other hand, the corresponding curvature reads
\beq\label{F0}
\bmF= d\bmA +\bmA\wedge \bmA
= \mathcal T^{\A} \bmP_{\A} +  \frac{1}{2} \mathcal R^{\A \B} \bmJ_{\A \B},
\eeq
where $\mathcal T^{\A}$ and $\mathcal R^{\A \B}$ are the torsion and the curvature forms
\beq\label{TandR}
\mathcal T^{\A} = d  E^{\A} + \Omega^{\A}_{\;\;\B}\wedge  E^{\B}
,\qquad
\mathcal R^{\A\B}= d \Omega^{\A\B}+\Omega^{\A}_{\;\; \C} \wedge \Omega^{\C\B}.
\eeq

We choose these fields to be such that, under a Poincar\'e transformation with gauge parameter $\varepsilon = \Upsilon^{\A} \bmP_{\A} + \frac12 \Theta^{\A\B} \bmJ_{\A\B}$, they transform only under the lower-dimensional Lorentz group $\mathfrak{so}(d,1)$ with parameters  $\Theta^{AB}$,  including as well diffeomorphisms with parameter $\Xi^{\hmu} $, we obtain the set of transformations
\beq\label{gaugetr}
\bal
&\delta E^{n}= \mathcal L_\Xi E^{n},
\\
&\delta  E^{A}= \mathcal L_\Xi E^{A} - \Theta^{A}_{\;\;B}  E^{B}\\
&\delta \Omega^{An}= \mathcal L_\Xi \Omega^{An}
-\Theta^A_{\;\;B} \Omega^{Bn},
\\
&\delta \Omega^{AB}= \mathcal L_\Xi \Omega^{AB}+ d\Theta^{AB}-\Omega^{[A}_{\;\;\,C}\Theta^{B]C} ,
\eal
\eeq 
where $\mathcal L$ stands for the Lie derivative. Actually, $\bmA$ is related to the $\mathfrak{iso} (d+1,1)$ gauge fields 
\beq\label{relconnectionLin}
\bmtA= \tilde E^{\A} \bmP_{\A} +\frac{1}{2} \tilde \Omega^{\A \B} \bmJ_{\A \B},
\eeq
by a gauge transformation with an element $e^{\Psi^{A} \bmJ_{A n}}e^{\Phi^{\A} \bmP_{\A}}$ belonging to the coset $\mathfrak{iso}(d+1,1)/\mathfrak{so}(d,1)$ (for details, see \cite{Ivanov:1981wn}),
\beq\label{nlA}
\bmA = e^{\Psi^{A} \bmJ_{A n}}e^{\Phi^{\A} \bmP_{\A}} \left(\bmtA +d\right) e^{-\Phi^{\B} \bmP_{\B}}e^{-\Psi^{B} \bmJ_{B n}}.
\eeq
In fact, the vielbein and spin connection transforms under the $\mathfrak{iso} (d+1,1)$ Poincar\'e transformations in the standard way, i.e.
\beq
\bal
&\delta \tilde E^{\A}= d \Upsilon^{\A} +\tilde \Omega^{\A}_{\;\;\B} \Upsilon^{\B} -\Theta^{\A}_{\;\;\B}\tilde E^{\A} ,
\\
&\delta \tilde \Omega^{\A\B} = d\Theta^{\A\B}-\tilde \Omega^{[\A}_{\;\;\,\C}\Theta^{\B]\C} .
\eal
\eeq
In addition, the St\"uckelberg fields $\Phi^{\A}$ and $\Psi^{A}$ obey the transformation rules  
\beq
\delta \Phi^{\A}= \Upsilon^{\A}-\Theta^{\A}_{\;\;\B} \Phi^{\B}, \quad \delta \Psi^{A}= \Theta^{A n}-\Theta^{A}_{\;\;B} \Psi^{B}.
\eeq
In terms of these fields, the non-linear gauge fields can be expressed as
\beq\label{EOmegaNLrel}
\bal
& E^n=E'^n
- E'^A\Psi_A
-\frac12E'^n\Psi^A\Psi_A+\frac{1}{3!}E'^A\Psi_A\Psi^B\Psi_B
+O\left(\Psi^4\right),
\\
&E^A=E'^A
+ E'^n\Psi^A
-\frac12 E'^B\Psi_B\Psi^A-\frac{1}{3!}E'^n\Psi^A\Psi^B\Psi_B
+O\left(\Psi^4\right),
\\
&\Omega^{An}= \tilde \Omega^{An}-d\Psi^A -\tilde\Omega^A_{\;\; B} \Psi^B
+\frac12\Psi_B\Psi^{[A}\tilde \Omega^{B]n}+\frac{1}{3!}\tilde \Omega^{A}_{\;\;\,B}\Psi^B\Psi^C \Psi_C
+O\left(\Psi^4\right),
\\
&\Omega^{AB}=\tilde \Omega^{AB}
-\Psi^{[A}\tilde \Omega^{B]n}+\frac12\Psi^{[A}\tilde \Omega^{B]}_{\;\;\,C}\Psi^C+\frac{1}{3!}\Psi^{[A}\tilde \Omega^{B]n} \Psi^C\Psi_C + O\left(\Psi^4\right).
\eal
\eeq
where we have defined the translational-invariant vielbein 
\beq\label{TinvE}
 E'^{\A}= \tilde  E^{\A}- d\Phi^{\A} -\tilde \Omega^{\A}_{\;\;\B}\Phi^{\B}.
\eeq
Notice that by construction the fields $ E^{\A}$ are invertible. In fact, we define the inverse vielbein $E_{\A}^{\;\;\hmu}$ satisfying $E_{\A}^{\;\;\hmu}E^{\A}_{\;\;\hnu}=\delta^{\hmu}_{\hnu}$ and $E^{\A}_{\;\;\hmu}  E_{\B}^{\;\;\hmu}=\delta^{\A}_{\B}$. In addition,  we can define the spacetime metric and its inverse.
\beq
G_{\hmu\hnu}=\eta_{\A \B}  E^{\A}_{\;\;\hmu} E^{\B}_{\;\;\hnu},\qquad 
G^{\hmu\hnu}= \eta^{\A\B}  E_{\A}^{\;\;\hmu} E_{\B}^{\;\;\hnu}.
\eeq
It is also convenient to introduce an affine connection $\Gamma^{
\hrho}_{\;\;\hmu\hnu}$ by means of the vielbein postulate
\beq\label{hdvielpost}
\hat \nabla_{\hmu} E^{\A}_{\;\;\hnu} = \partial_{\hmu} E^{\A}_{\;\;\hnu}
+ \Omega^{\A}_{\;\;\B \hmu} E^{\B}_{\;\;\hnu}- \Gamma^{
\hrho}_{\;\;\hmu\hnu} E^{\A}_{\;\;\hrho}=0.
\eeq
In terms of which the components of the torsion and the curvature read
\beq
\bal
&\mathcal T^{\hrho}_{\;\;\hmu\hnu}=E_{\A}^{\;\;\hrho}\mathcal  T^{\A}_{\;\;\hmu\hnu} =\Gamma^{\hrho}_{\;\;\hmu\hnu}-\Gamma^{
\hrho}_{\;\;\hnu\hmu}
\\
&\mathcal R^{\hrho}_{\;\;\hsigma\hmu\hnu}= E_{\A}^{\;\;\hrho}  E_{\B\hsigma} \mathcal R^{\A\B}_{\;\;\;\;\hmu\hnu}
=\partial_{\hmu} \Gamma^{\hrho}_{\;\;\hnu\hsigma}
-\partial_{\hnu} \Gamma^{\hrho}_{\;\;\hmu\hsigma}
+\Gamma^{\hrho}_{\;\;\hmu\hgamma} \Gamma^{\hgamma}_{\;\;\hnu\hsigma}
-\Gamma^{\hrho}_{\;\;\hnu\hgamma} \Gamma^{\hgamma}_{\;\;\hmu\hsigma}.
\eal
\eeq
Finally, with these ingredients we define the volume form 
\begin{equation}
 ^*1 = \frac{1}{(d+2)!}\epsilon_{\hat A_0\ldots \hat A_{d+1}} E^{\hat A_0}\wedge\ldots\wedge E^{\hat A_{d+1}},
\end{equation}
and a Hodge dual operation that we can use to  define the  Yang-Mills-like action
\beq\label{actionRR}
\bal
S&=- \frac12 \int \left\langle *\bmF \wedge \bmF  \right\rangle,
\eal
\eeq
where $\left\langle\cdots\right\rangle$ stands for an invariant metric on the $\mathfrak{iso}(d+1,1)$ algebra in $d+2$ dimensions. However, note that the Higgsing of the theory allows us to define a bi-linear form only invariant under the action of the  unbroken subgroup  $SO(d,1)$. With that criteria, the most general choice is
\beq\label{itso}
\bal
&\left\langle \bmJ_{AB} \bmJ_{CD}\right\rangle= \halpha_0 \left(\eta_{AC} \eta_{BD} -\eta_{AD} \eta_{BC}\right)
,&\qquad&
\left\langle \bmP_n \bmP_n\right\rangle= \halpha_1,
\\
&\left\langle \bmJ_{An} \bmJ_{Bn}\right\rangle= \halpha_2 \eta_{AB} ,
&\qquad&
\left\langle \bmP_{A} \bmP_{B}\right\rangle= \halpha_3 \eta_{AB} .
\eal
\eeq

\subsection{Aristotelian contraction}
As previously pointed out our goal is to implement the contraction \eqref{genresc} of the Poincar\'e algebra on the gravitational theory described by Eq. \eqref{actionRR}. This can be achieved after rescaling the gauge fields as
\begin{multicols}{2}
\begin{subequations} \label{splitfields}
\setlength{\abovedisplayskip}{-14pt}
\allowdisplaybreaks
\begin{align}
&\Omega^{ab}=\omega^{ab},\\
&\Omega^{an} = \sigma \,v^a, \\
&\Omega^{0a}=\varepsilon \,\mu^a,\\
&\Omega^{0n} = \varepsilon\sigma \,u,\\
&E^0 =\tau,\\
&E^a =e^a,\\
&E^n= \sigma\,\rho.
\\ \nonumber
\end{align}
\end{subequations}
\end{multicols}\noindent

Actually, in the limit $\varepsilon\rightarrow 0$, the connection shown in Eq. \eqref{relconnection} reduces to a non-Lorentzian one taking values on the elements of the algebra defined by \eqref{dipolealgext} and \eqref{Ideal}, which can be written as
\beq\label{decA}
\bmA = \bA+\bB,
\eeq
where
\beq\label{NLconnection}
\bA= \tau \,  \bH+ e^a  \bP_a +\frac{1}{2} \omega^{ab}  \bJ_{ab} + v^a  \bQ_a +  \rho \,\bQ ,\qquad \bB=  \mu^a \bG_a + u \bK .
\eeq
The corresponding contracted curvature is given by
\beq
\bmF= \bF + \mathcal D_{\bA} \bB,
\eeq
where $\bF=d\bA+\bA\wedge\bA$ is the curvature associated to the connection $\bA$, which reads
\beq\label{curvatureNL}
\bF = d\tau \, \bH+ \left(T^a+\sigma^2 v^a\wedge \rho\right)  \bP_a +\frac{1}{2} \left(R^{ab}-\sigma^2 v^a \wedge v^b \right)  \bJ_{ab}+ F^a \, \bQ_a +  f \,\bQ ,
\eeq
with
\begin{multicols}{2}
\begin{subequations} \label{curvNL}
\setlength{\abovedisplayskip}{-14pt}
\allowdisplaybreaks
\begin{align}
&T^a=De^a ,\\
&R^{ab}= d\omega^{ab} + \omega^a_{\;\;c} \wedge \omega^{cb}, \\
&F^a= Dv^a, \\
&f= d\rho+ e^a \wedge v_a,
\end{align}
\end{subequations}
\end{multicols}\noindent
and $D$ is the covariant exterior derivative with respect to $\omega^{ab}$. The term $\mathcal D_{\bA}\bB$, on the other hand, is the covariant derivative of $\bB$ with respect to $\bA$.
\beq
\mathcal D_{\bA}\bB= d\bB+[\bA,\bB]=\left(D\mu^a+\sigma^2 v^a \wedge u\right) \bG^a + \left(d u +\mu^a \wedge v_a\right)\bK.
\eeq
Notice also that in the limit $\varepsilon\rightarrow 0$, the non-vanishing components of the invariant tensor \eqref{itso} are
\beq\label{itsoLim}
\bal
&\left\langle \bJ_{ab} \bJ_{cd}\right\rangle= \halpha_0 \left(\delta_{ac} \delta_{bd} -\delta_{ad} \delta_{bc}\right)
,&\qquad&
\left\langle \bQ_a \bQ_{b}\right\rangle= \sigma^2 \halpha_2 \delta_{ab} ,
&\qquad&
\left\langle \bQ \bQ\right\rangle= \sigma^2\halpha_1
\\
&\left\langle \bP_{a} \bP_{b}\right\rangle= \halpha_3 \delta_{ab} 
,&\qquad&
\left\langle \bH \bH\right\rangle= -\halpha_3
,
&\qquad&
\eal
\eeq
whereas the components $\left\langle \bG_{a} \bG_{b}\right\rangle$ and $\left\langle \bK \bK \right\rangle$ vanish. This means that the gauge fields $\mu^a$ and $u$ entering $\bB$ decouple from the rest of the gauge fields in the Aristotelian limit and do not appear in the action \eqref{actionRR} after the contraction. Thus, for simplicity we remove the connection $\bB$ from the analysis and consider only the connection $\bA$.

The transformations in Eqs. \eqref{gaugetr} lead to the following symmetry transformations for the gauge fields in Eqs. \eqref{NLconnection} when $\varepsilon\rightarrow 0$
\begin{multicols}{2}
\begin{subequations} \label{gaugetrNLlinear}
\setlength{\abovedisplayskip}{-14pt}
\allowdisplaybreaks
\begin{align}
&\delta \tau =\mathcal L_\Xi  \tau  ,
\\
&\delta e^a=\mathcal L_\Xi  e^a - \theta^a_{\;\;b}e^b,
\\
&\delta \rho=\mathcal L_\Xi  \rho,
\\
&\delta v^a=\mathcal L_\Xi  v^a - \theta^a_{\;\;b}v^b,
\\
&\delta \omega^{ab}=\mathcal L_\Xi \omega^{ab}  +  D\theta^{ab},
\\ \nonumber
\end{align}
\end{subequations}
\end{multicols}\noindent
where the gauge parameters are related to the ones in Eqs. \eqref{gaugetr} by $\Theta^{ab}=\theta^{ab}$. From the contracted algebra perspective these transformations can be obtained from a gauge transformation of the connection defined in Eq. \eqref{NLconnection} as, $\delta \bA = \mathcal L_\xi \bA+d\blambda+[\bA,\blambda]$, with $\blambda=  \frac12 \theta^{ab} \bJ_{ab}$. 

In the Aristotelian limit $\varepsilon\rightarrow0$, the linear connection \eqref{relconnectionLin} can also be decomposed in the form \eqref{decA} with
\beq
\tilde \bA= \tilde \tau \,  \bH+ \tilde e^a  \bP_a +\frac{1}{2} \tilde \omega^{ab}  \bJ_{ab} + \tilde v^a  \bQ_a + \tilde  \rho \,\bQ .
\eeq
Using \eqref{TinvE} we can define the translational-invariant vielbein
\beq\label{Tinvtaue}
\bal
\lambda &= \tilde \tau- d\phi 
\\
h^a&= \tilde  e^a- D\phi^a -\sigma^2 v^a \varphi,
\\
a &= \tilde \rho- d\varphi +v_a \phi^a,
\eal
\eeq
where we have defined the following rescaling for the field $\Phi^{\A}$
\beq
\Phi^n= \sigma \varphi,\quad
\Phi^0=\phi,\quad
\Phi^a=\phi^a.
\eeq
Similarly, using \eqref{EOmegaNLrel} and rescaling $\Psi^A$ in the form
\beq
\Psi^0=\varepsilon\sigma \psi,\qquad
\Psi^a=\sigma \psi^a,
\eeq
the Aristotelian non-linear fields can be expressed in terms of the linear ones as
\beq\label{EOmegaNL}
\bal
&\tau =\lambda
\\
&e^a =h^a
+ \sigma^2 a \psi^a
-\frac{\sigma^2}{2} h^b\psi_b\psi^a-\frac{\sigma^4}{3!} a \psi^a\psi^b\psi_b
+O\left(\psi^4\right),
\\
& \rho =a
- h^a\psi_a
-\frac{\sigma^2}{2} a \psi^a\psi_a+\frac{\sigma^2}{3!}h^a\psi_a\psi^b\psi_b
+O\left(\psi^4\right),
\\
&v^a= \tilde v^a- \tilde D\psi^a
+\frac{\sigma^2}{2}\psi_b\psi^{[a}\tilde v^{b]}
+\frac{\sigma^2}{3!}\tilde \omega^{a}_{\;\;\,b}\psi^b \psi^c\psi_c
+O\left(\psi^4\right),
\\
&\omega^{ab}=\tilde \omega^{ab}
-\sigma^2\psi^{[a}\tilde v^{b]}+\frac{\sigma^2}{2}\psi^{[a}\tilde \omega^{b]}_{\;\;\,c}\psi^c+\frac{\sigma^4}{3!}\psi^{[a}\tilde v^{b]}\psi^c\psi_c + O\left(\psi^4\right),
\\
\eal
\eeq
where fields $\lambda$, $h^a$, $a$, $\tilde v^a$ and $\tilde\omega^{ab}$ transform as
\begin{multicols}{2}
\begin{subequations} 
\setlength{\abovedisplayskip}{-14pt}
\allowdisplaybreaks
\begin{align}
&\delta \lambda =\mathcal L_\Xi  \lambda ,
\\
&\delta h^a=\mathcal L_\Xi  h^a - \theta^a_{\;\;b}h^b-\sigma^2b^a a ,
\\
&\delta a=\mathcal L_\Xi  a + b_a h^a ,
\\
&\delta \tilde v^a=\mathcal L_\Xi \tilde v^a +  \tilde Db^a - \theta^a_{\;\;b}\tilde v^b,
\\
&\delta \tilde \omega^{ab}=\mathcal L_\Xi \tilde \omega^{ab}  +  \tilde D\theta^{ab}-\sigma^2 \tilde v^{[a}b^{b]},
\\ 
&\delta \psi^a = \mathcal L_\Xi  \psi^a + b^a -\theta^a_{\;\;b}\psi^b.
\end{align}
\end{subequations}
\end{multicols}\noindent
Notice that, apart from the parameter associated to local rotations, we have defined $\Theta^{an}=\sigma\, b^a$.
Since we are interested in expressing the action \eqref{actionRR} in terms of the rescaled fields given in Eq. \eqref{splitfields}, we notice that the $d+2$ metric and its inverse can be decomposed as
\beq
G_{\hmu\hnu}= g_{\hmu\hnu}+\sigma^2\rho_{\hmu} \rho_{\hnu},\qquad
G^{\hmu\hnu}= g^{\hmu\hnu} +\frac{1}{\sigma^2}\rho^{\hmu} \rho^{\hnu},
\eeq
where the inverse vielbein has been rescaled as $E_{\A}^{\;\;\hmu}=(\tau^{\hmu}, e_a^{\;\,\hmu},\sigma^{-1}\rho^{\hmu})$ and we have defined
\beq
g_{\hmu\hnu}=-\tau_{\hmu}\tau_{\hnu} + \delta_{ab}e^a_{\;\,\hmu} e^b_{\;\,\hmu} 
,\qquad
g^{\hmu\hnu}=-\tau^{\hmu} \tau^{\hnu} + \delta^{ab}e_a^{\;\,\hmu} e_b^{\;\,\hnu}.
\eeq
The fields satisfy the orthogonality relations
\beq
 \rho^{\hmu} g_ {\hmu\hnu}=0=\rho_{\hmu} g^{\hmu\hnu}
,\qquad 
\rho^{\hmu}\rho_{\hmu}=1
,\qquad
g^{\hmu\hrho}g_{\hrho\hnu}+ \rho^{\hmu}\rho_{\hnu}=\delta^{\hmu}_{\hnu}.
\eeq
and the square root metric determinant can be written as
\beq
\sqrt{|G|}
=\sigma \sqrt{|g\rho|}, \qquad |g\rho| =\frac{1}{(d+1)!}\epsilon_{a_0 \cdots a_d} \epsilon^{\mu_0\cdots \mu_{d+1}} e^{a_0}_{\;\,\mu_0}\cdots e^{a_d}_{\;\,\mu_d} \rho_{\mu_{d+1}}.
\eeq
All this allows us to expand the higher-dimensional action in powers of $\sigma$, 
\beq
S= \frac{1}{\sigma}S_0 +\sigma S_1+\sigma^3 S_2 + O(\sigma^5),
\eeq
with the leading order terms taking the form
\beq\label{S012}
\bal
S_0&=- \int d^{d+2}x \, \sqrt{|g\rho|}\, 
g^{\hmu\hrho}\rho^{\hnu} \rho^{\hsigma}
\left[ \frac{\halpha_0}{2} R^{ab}_{\;\,\;\,\hmu\nu} \,R_{ab\hrho\hsigma}
+ \halpha_3 T^a_{\;\,\hmu\hnu} \,T_{a\hrho\hsigma}-\halpha_3 \partial_{[\hmu}\tau_{\hnu]}
\partial_{[\hrho}\tau_{\hsigma]}\right],
\\[8pt]
S_1&= - \int d^{d+2}x\, \sqrt{|g\rho|}\, 
\bigg[ g^{\hmu\hrho}g^{\hnu\hsigma}\left(
\frac{\halpha_0 }{4}  R^{ab}_{\;\,\;\,\hmu\nu} \,R_{ab\hrho\hsigma}
+\frac{\halpha_3}{2} 
T^a_{\;\,\hmu\hnu} \,T_{a\hrho\hsigma}
-\frac{\halpha_3}{2}\partial_{[\hmu}\tau_{\hnu]}
\partial_{[\hrho}\tau_{\hsigma]}
\right)
\\
&\hskip.8truecm
+2\halpha_3 g^{\hmu\hrho} \rho^{\hnu}  T_{a\hmu\hnu}\; v^a_{\;\hrho}
+g^{\hmu\hrho} \rho^{\hnu} \rho^{\hsigma} \left( \halpha_2 F^a_{\;\,\hmu\hnu}\,F_{a\hrho\hsigma}
- \halpha_0 R_{ab\hmu\hnu}\; v^a_{\;[\hrho}v^b_{\;\,\hsigma]}
+\halpha_1 f_{\hmu\hnu}\,f_{\hrho\hsigma}
\right) 
\bigg],
\\[8pt]
S_2&= -\frac12 \int d^{d+2}x\, \sqrt{|g\rho|}\, 
 \Bigg[
 g^{\hmu\hrho}g^{\hnu\hsigma}
\left(\halpha_2 F_{\;\,\hmu\hnu}^{a}\,F_{a \hrho\hsigma}
-
\halpha_0 R_{ab\hmu\hnu}\; v^a_{\;[\hrho}v^b_{\;\,\hsigma]} 
+
\halpha_1 
f_{{\hmu}{\hnu}}\,f_{ \hrho\hsigma}\right)
\\
&\hskip3.5truecm
+  \halpha_0 g^{\hmu\hrho} \rho^{\hnu} \rho^{\hsigma}
\;v_{a[\hmu}v_{b|\hnu]}\; v^a_{\;[\hrho}v^b_{\;\,\hsigma]}
+ 2\halpha_3 g^{\hmu\hrho} 
\;v_{a\hmu}\; v^a_{\;\,\hrho}
\Bigg].
\eal
\eeq

\subsection{Dimensional reduction}
\label{sec:dimred}

Before starting with the dimensional reduction procedure, we would like to point out that, as it generically happens with gravitational theories,  local translations are "spontaneously broken" \cite{ivanov1982gauge}. Nonetheless, we are interested in a system where the fracton charge (momentum in the extra direction) is conserved. Therefore, we will require the existence of a spacelike (transverse) killing vector $\mathcal K$, and to use coordinates $x^{\hmu}=(x^\mu,z)$ such that $\mathcal K=\partial_z$. Moreover, from the perspective of this construction the fracton $U(1)$ transformations can be understood as translations along the extra spacetime dimension (see Eq. \eqref{genresc}). Therefore, since the generator $\mathbf Q$ commutes with all the rest after the contraction it is natural to require the existence of a Killing vector $\mathcal K$. That would guarantee that gauge fields remain invariant when a transverse diffeo with constant parameter is applied. Thus we require
\beq\label{dzA}
    \mathcal L_{\mathcal K}\mathbf A = \partial_z \mathbf A = 0,
\eeq
which implies all components of the fields to be $z$-independent. 
After doing so, we introduce the gauge fixing conditions 
\beq\label{gf1}
\rho_z=1,
\eeq
From the orthogonality relations between the higher-dimensional vielbein and its inverse, it follows that setting \eqref{gf1} implies
\beq\label{gf2}
\rho^{\hmu} =\delta^{\hmu}_z,\quad 
\tau_z=0=
e^a_{\;\,z}
,\quad
\tau^z=- \rho_\mu  \tau^\mu ,\quad e^z_{\;\,a} = - \rho_\mu   e_{\;\,a}^\mu,
\eeq
whereas the rest of the fields satisfy the relations
\begin{multicols}{2}
\begin{subequations} \label{relsnonrel}
\setlength{\abovedisplayskip}{-14pt}
\allowdisplaybreaks
\begin{align}
&\tau^{\mu} \tau_{\mu} = 1,\\
&e^a_{\;\,\mu} \tau^{\mu} =0,\\
&\tau_{\mu} e_a^{\;\,\mu} =0,\\
&e_a^{\;\,\mu} e^b_{\;\,\mu} = \delta^b_a,\\
&\tau^{\mu} \tau_{\nu}+e_a^{\;\,\mu} e^a_{\;\,\nu} = \delta^{\mu}_{\nu} ,\\
\nonumber
\end{align}
\end{subequations}
\end{multicols}\noindent
familiar from non-Lorentzian geometry. The higher-dimensional metric tensor then takes the form
\beq\label{gaugefixedG}
G_{\hmu\hnu} dx^{\hmu} dx^{\hnu} = g_{\mu\nu} dx^\mu dx^\nu + \sigma^2 (dz + \rho_\mu dx^\mu) (dz + \rho_\nu dx^\nu),
\eeq
and the determinant of the metric reduces to
\beq
\sqrt{|g\rho|}
=\sigma \sqrt{|g|.}
\eeq

In order to make the dipole symmetry explicit, we express the non-linear fields $\rho_\mu$ and $v_{\hmu}$ in terms of $a_\mu$, $\tilde v^a_{\;\,\hmu}$, $\psi^a$, $\omega^{ab}_{\;\,\;\,\hmu}$, $\tau_\mu$ and $e^a_{\;\,\mu}$. At leading order in $\sigma$ we can write
\beq
\bal
&\rho_\mu = a_\mu - e^a_{\;\,\mu} \psi_a + O(\sigma^2),
\\
&v^a _{\;\,\hmu}= \tilde v^a _{\;\,\hmu} -D_{\hmu}\psi^a + O(\sigma^2).
\eal
\eeq
Thus, by defining
\beq
\tilde F^a= D\tilde v^a, \quad
\tilde f = da+e^a \tilde v_a,
\eeq
the curvatures $F^a_{\;\,\hmu\hnu}$ and $f_{\hmu\hnu}$ can be written as
\beq
\bal
&F^a_{\;\,\hmu\hnu} = \tilde F^a_{\;\,\hmu\hnu}-R^{ab}_{\;\,\;\,\hmu\hnu} \psi_b + O(\sigma^2),
\\
&f_{\hmu\hnu}= \tilde f_{\hmu\hnu}
- T^a_{\;\,\hmu\hnu}\psi_a
+ O(\sigma^2).
\eal
\eeq
Finally, as a last gauge fixing we impose
\beq\label{omegazandvz}
\omega^{ab}_{\;\,\;\,z}=0,\qquad \tilde v^a_{\;\,z}=0.
\eeq
Implementing all these conditions, the transformations of the $(d+1)$-dimensional fields take the form
\begin{multicols}{2}
\begin{subequations} \label{gaugeTNL}
\setlength{\abovedisplayskip}{-14pt}
\allowdisplaybreaks
\begin{align}
&\delta \tau_\mu= \mathfrak L_\xi \tau_\mu ,\\
&\delta e^a_{\;\,\mu}=  \mathfrak L_\xi  e^a_{\;\,\mu} 
- \theta^a_{\;\; b} e^b_{\;\,\mu} ,\\
&\delta a_\mu= \mathfrak L_\xi  a_\mu - \partial_\mu \epsilon+ b_a e^a_{\;\,\mu} ,\\
&\delta \tilde v^a_{\;\,\mu}= \mathfrak L_\xi   \tilde v^a_{\;\,\mu}  + D_\mu b^a - \theta^a_{\;\;b}v^b_{\;\,\mu} ,\\
&\delta \omega^{ab}_{\;\,\;\,\mu}= \mathfrak L_\xi  \omega^{ab}_{\;\,\;\,\mu} +  D_\mu\theta^{ab} ,\\
&\delta \psi^a=\mathfrak L_\xi\psi^a +b^a-\theta^a_{\;\;b} \psi^b,
\end{align}
\end{subequations}
\end{multicols}\noindent
where $\mathfrak L$ denotes the Lie derivative in $d+1$ dimensions and the higher-dimensional diffeomorphism parameter has been redefined as 
\beq
\Xi=\Xi^{\hmu}\partial_{\hmu}=\xi^\mu \partial_\mu- \epsilon \partial_z = \xi- \epsilon \partial_z.
\eeq
where we have remaned $\Xi^n=-\epsilon$. Demaning that the gauge conditions \eqref{gf1} and \eqref{gf2} are invariant under gauge transformations \eqref{gaugeTNL}, restricts the diffeomorphism parameter $\xi^\mu$ and the gauge paramenters $\theta^{ab}$ and $b^a$ to be $z$-independent.

Due to the conditions \eqref{gf1},\eqref{gf2} and \eqref{omegazandvz}, the gauge connection \eqref{NLconnection} satisfies $\bA_z=\bQ$. This together with the fact that the fields are $z$-independent imply that the field strength two-form \eqref{curvatureNL} satisfies 
\beq
\bF_{z\mu}=\partial_{[z}\tau_{\mu]} \bH + T^a_{\;\,z\mu} \bP_a + \frac12 R^{ab}_{\;\,\;\,z\mu}\bJ_{ab}+ F^a_{\;\,z\mu}\bQ^a+ f_{z\mu}\bQ=0.
\eeq
As a consequence, the action $S_0$ in \eqref{S012} vanishes. The condition \eqref{dzA} allows to trivially integrate over the coordinate $z$ in the action. Introducing the new constant
\beq
\alpha_n =\sigma \halpha_n \int dz.
\eeq
we find the following gauge-fixed action
\beq\label{S1S2}
S= S_1 + \sigma^2 S_2 + O(\sigma^4),
\eeq
where
\beq
\bal
&S_1=-\frac12 \int d^{d+1}x\, \sqrt{|g|}\, 
\left[ \frac{\alpha_0}{2}R^{ab}_{\;\,\;\,\mu\nu} \,R_{ab}^{\;\,\;\,\mu\nu}
+\alpha_3 T^a_{\;\,\mu\nu} T_a^{\;\,\mu\nu}
-\alpha_3 \partial_{[\mu}\tau_{\nu]}
\partial^{[\mu}\tau^{\nu]}
\right],
\\[8pt]
&S_2= -\frac12  \int d^{d+1}x\, \sqrt{|g|}\, 
\bigg[\alpha_2 \left(\tilde F^a_{\;\,\mu\nu}-R^{ab}_{\;\,\;\,\mu\nu}\psi_b \right)\left(\tilde F_{a}^{\;\,\mu\nu}-R_{ac}^{\;\,\;\,\mu\nu}\psi^c \right)
\\&
-\alpha_0
R^{ab\mu\nu}\left(\tilde v_{a\mu}-D_\mu\psi_a\right) \left(\tilde v_{b\nu} -D_\nu \psi_b\right)
+\alpha_1 \left(\tilde f_{\mu\nu}-T^a_{\;\,\mu\nu}\psi_a\right) \left(\tilde f^{\mu\nu}-T^{b\mu\nu}\psi_b \right)
\\&
+ 2 \alpha_3 \left(\tilde v^a_{\;\,\mu}-D_\mu \psi^a\right)\left(\tilde v_a^{\;\,\mu}-D^\mu \psi_a\right)
\bigg].
\eal
\eeq
Notice that, in the strict limit $\sigma=0$, the limit pocedure leads to $S=S_1$. However, other limit can be defined that lead to the action $S_2$ instead. In the following we discuss a few interesting cases:

\vspace{.2cm}\noindent$\bullet\;$\textbf{Pretko's theory} - we can introduce the rescaling $\alpha_n\rightarrow \frac{1}{\sigma^2}\alpha_n$ and introduce auxiliary fields $\lambda^{ab}_{\;\,\;\,\mu\nu}$, $\lambda^a_{\;\,\mu\nu}$ and $\lambda_{\mu\nu}$, which allow to rewrite \eqref{S1S2} as
\beq
\bal
S&=  \alpha_0\int d^{d+1}x\, \sqrt{|g|}\, 
\left(\frac{\sigma^2}{2}
\; \lambda^{ab}_{\;\,\;\,\mu\nu} \,\lambda_{ab}^{\;\,\;\,\mu\nu}
-
\; \lambda_{ab}^{\;\,\;\,\mu\nu} \,R^{ab}_{\;\,\;\,\mu\nu}
\right)
\\&
+ \alpha_3\int d^{d+1}x\, \sqrt{|g|}\, 
\left(\sigma^2
\; \lambda^a_{\;\,\mu\nu} \,\lambda_{a}^{\;\,\mu\nu}
-2
\; \lambda_a^{\;\,\mu\nu} \,T^a_{\;\,\mu\nu}
-\sigma^2 \lambda_{\mu\nu}\lambda^{\mu\nu}-2\lambda^{\mu\nu}\partial_{[\mu}\tau_{\nu]}
\right)
+ S_2 + O(\sigma^2).
\eal
\eeq
Indeed, one can see that integrating out the auxiliary fields yields the action \eqref{S1S2} after properly rescaling the constants $\alpha_n$. Now, in the limit $\sigma\rightarrow0$, $\lambda^{ab}_{\;\,\;\,\mu\nu}$, $\lambda^a_{\;\,\mu\nu}$ and $\lambda_{\mu\nu}$ become Lagrange multipliers enforcing the constraints
\beq
R^{ab}_{\;\,\;\,\mu\nu}=0,\qquad
T^a_{\;\,\;\,\mu\nu}=0,\qquad
\partial_{[\mu}\tau_{\nu]}=0,
\eeq
and thus leading to the action $S=S_2$ on flat space. For $\alpha_3 =0$ and $\alpha_2=1$, the action boils down to
\beq\label{actionflat}
S=-\frac12 \int d^{d+1}x \left[ \tilde F^a_{\;\,\mu\nu}\tilde F_a^{\;\,\mu\nu}+\alpha_1 \tilde f_{\mu\nu} \tilde f^{\mu\nu}\right].
\eeq
Splitting the index $\mu$ into space and time components $\mu=(0,i)$, choosing $\tau_\mu=(1,0,\dots,0)$, $e^a_{\;\,\mu} =(0, \delta^a_i$), $\omega^{ab}_{\;\,\;\,\mu}=0$, and gauge fixing $a_\mu = (\phi,0,\dots,0)$ we can write
\beq
\tilde F^a_{\mu\nu} = \partial_{[\mu} \tilde v^a_{\;\,\nu]},\qquad \tilde f_{0i}=-\partial_i \phi-\tilde v_{i0},
\qquad \tilde f_{ij}= \tilde v_{[ij]},
\eeq
where we have defined $\tilde v_{\mu\nu}= e_{a\mu}\tilde 
 v^a_{\;\,\nu}$. We now split the field $\tilde v^a_{\;\,\mu}$  as
\beq\label{decompv}
\tilde v^a_{\;\,\mu} = A^a_{\;\,\mu}+B^a_{\;\,\mu},
\eeq
where $u^a_{\;\,\mu}$ is the part of $\tilde v^a_{\;\,\mu}$ that solves $\tilde f_{\mu\nu}=0$ and thus has the form
\beq
A_{0i} = - \partial_i  \phi
,\quad
A_{ij} = A_{ji}.
\eeq
Notice that the curvature $f_{\mu\nu}$ is gauge invariant in the absence of torsion and the second term in \eqref{actionflat} is now a mass term for the fields $B_{0i}$ and $B_{ij}$. In the low energy regime of the theory we can keep only the gapless modes and therefore neglect $B^a_{\;\,\mu}$. In this case, defining 
\beq
F_{\mu\nu k} \equiv \delta_{ak} \tilde F^a_{\;\,\mu\nu}
\eeq
leads to the action \eqref{ActionPretko} proposed in \cite{Pretko:2016lgv}. The gauge transformation for $\phi$ and $A_{ij}$ follow from \eqref{gaugeTNL} and the gauge invariance of the gauge condition $a_i=0$. This yields\footnote{Notice that this result also holds in curved space, where setting $a_\mu e_a^{\;\,\mu} =0$ as a gauge condition yields $b_a=e_a^{\;\,\mu}\partial_\mu\epsilon$, which leads to the transformation law \eqref{transformationAamu} for $A^a_{\;\,\mu}$.}
\beq\label{relbandetaflat}
b_a =\delta^i_a \partial_i \epsilon
\quad\Rightarrow\quad
\delta\phi=-\dot \epsilon
,\quad
\delta A_{ij}=\partial_i\partial_j \epsilon  ,
\eeq
which matches \eqref{gtflat}.

\vspace{.2cm}\noindent$\bullet\;$\textbf{Proca extension} - Turning on the constant $\alpha_3$ in the previous example leads to a Proca extension of Fracton electrodynamics. Indeed, by renaming, and $\alpha_3 =m^2$ 
, the action \eqref{ActionPretko} reads
\beq
S = - \int d^{d+1}x\, 
\bigg[ \frac12 D_{[\mu}A^a_{\;\,\nu]}D^{[\mu}A_a^{\;\,\nu]}
+  m^2 \left( A^a_{\;\,\mu} - \partial_\mu \psi^a\right)\left( A_a^{\;\,\mu} - \partial^\mu \psi_a\right) 
\bigg] +\ldots,
\eeq
where $\ldots$ in the action contains the kinetic term for the $B$ field, and the  interactions between the gauge field $A^a_{\;\,\mu}$ and $B^a_{\;\,\mu}$. Notice again that the $B$ field will be massive and gauge invariant.

\vspace{.2cm}\noindent$\bullet\;$\textbf{Curved space generalization} - Implementing the rescaling $\alpha_1\rightarrow \sigma^{-2}\alpha_1$ and $\alpha_2\rightarrow \sigma^{-2}\alpha_2$ and taking the limit $\sigma\rightarrow 0$ in the action \eqref{S1S2} we find
\beq
\bal
S&=-\frac12 \int d^{d+1}x\, \sqrt{|g|}\, 
\bigg[ \frac{\alpha_0}{2}R^{ab}_{\;\,\;\,\mu\nu} \,R_{ab}^{\;\,\;\,\mu\nu}
+\alpha_3 T^a_{\;\,\mu\nu} T_a^{\;\,\mu\nu}
-\alpha_3 \partial_{[\mu}\tau_{\nu]}
\partial^{[\mu}\tau^{\nu]}
\\
&
+\alpha_2 \left(\tilde F^a_{\;\,\mu\nu}-R^{ab}_{\;\,\;\,\mu\nu}\psi_b \right)\left(\tilde F_{a}^{\;\,\mu\nu}-R_{ac}^{\;\,\;\,\mu\nu}\psi^c \right)
+\alpha_1 \left(\tilde f_{\mu\nu}-T^a_{\;\,\mu\nu}\psi_a\right) \left(\tilde f^{\mu\nu}-T^{b\mu\nu}\psi_b \right)
\bigg],
\eal
\eeq
which corresponds to a torsion-full generalization of the action proposed in \cite{Pena-Benitez:2021ipo}. Indeed, the term proportional to $\alpha_2$ is precisely the action \eqref{Actioncurvedimp2}.

\vspace{.2cm}\noindent$\bullet\;$\textbf{Constrained curved background} - Another possibility is to treat the action \eqref{S1S2} perturbatively in the parameter $\sigma$, by assuming the fields $\tilde v^a_{\;\,\mu}$ and $a_{\mu}$ do not backreact on the geometry and considering the gravitational fields to be on-shell 
. We consider the case $\alpha_3=0$ and $\alpha_0/2=\alpha_2=1$. The action $S_1$ then reduces to an Aristotelian version of the Stephenson-Kilmister-Yang model  \cite{stephenson1958quadratic,kilmister1961use,Yang:1974kj}. The field equations coming from $S_1$ after varying with respect to the metric fields and the spin connection read
\begin{align}
& \delta\omega^{ab}_{\;\,\;\,\mu}:&\quad& 
D^\mu  R_{ab\mu\nu} =0, \label{feq1}
\\
&\delta h^{\alpha\beta}:&\quad&
\frac12 R^{ab}_{\;\,\;\,\mu(\alpha}R_{ab\;\,\;\beta)}^{\;\,\;\,\mu}
-\frac{1}{4}
g_{\alpha\beta} 
R^{ab}_{\;\,\;\,\mu\nu} R_{ab}^{\;\,\;\,\mu\nu}=0,
\label{feq2}
\\
&\delta \tau^\alpha: &\quad&
\tau_\alpha R^{ab}_{\;\,\;\,\mu\nu} R_{ab}^{\;\,\;\,\mu\nu}
-2\tau^\mu R^{ab}_{\;\,\;\,\mu\nu}R_{ab\alpha}^{\;\,\;\,\;\;\nu}=0.
\label{feq3}
\end{align}
We consider a solution with vanishing torsion $T^a_{\mu\nu}=0=\partial_{[\mu}\tau_{\nu]}$. One consequence of this is that the resulting action $S_2$ is gauge invariant even in absence of the St\"uckelbeg field $\psi^a$
thanks to the field equation \eqref{feq1}. This means that on such gravitational backgrounds it is possible to describe a phase of the system with unbroken dipole symmetry where $\psi^a=0$. Moreover, as in the flat case, after decomposing $\tilde v^a_{\;\,\mu}$ in the form \eqref{decompv}, the term $\tilde f_{\mu\nu}\tilde f^{\mu\nu}$ is a mass for $B^a_{\;\,\mu}$ and therefore this field can be neglected in a low energy description of the system. Under all these considerations, the action $S_2$ takes the simple form
\beq\label{S2RAA}
S_2= -  \int d^{d+1}x\, \sqrt{|g|}\, 
\bigg[\frac12 \tilde F^a_{\;\,\mu\nu}
\tilde F_{a}^{\;\,\mu\nu}
-
R^{ab\mu\nu} A_{a\mu} A_{b\nu}
\bigg].
\eeq
This is precisely the action \eqref{Actioncurvedimp} when the gauge fixing condition
\beq\label{gfa}
a_\mu = \phi \,\tau_\mu
\eeq
is imposed. Indeed, the generalization of $A^a_{\;\,\mu}$ in the curved space case has the form
\beq\label{solAamucurved}
A^a_{\;\,\mu} = \frac{1}{2}  \partial_{[\alpha}  a_{\nu]}
\left(\delta^\alpha_\mu +\tau^\alpha  \tau_\mu \right)e^{a\nu} + A_{ab} e^b_{\;\,\mu}
,\quad A_{ab}=A_{ba}.
\eeq
When the conditions \eqref{dzA}, \eqref{gf1}, \eqref{gf2} and \eqref{omegazandvz} are imposed, at leading order in the $\sigma$-expansion the vielbein postulate \eqref{hdvielpost} and its inverse relation lead to the lower-dimensional vielbein postulate and its inverse
\beq
\nabla_\mu \tau_\nu =0, \quad \nabla_\mu e^a_{\;\,\nu}=0,\quad
\nabla_\mu \tau^\nu =0, \quad \nabla_\mu e_a^{\;\,\nu}=0.
\eeq
where $\nabla_\mu$ acts with $\omega^{ab}_{\;\,\;\,\mu}$ on tangent space indices and with $\Gamma^\sigma_{\mu\nu}$ on spacetime indices. Using these relations we can write the projections of the gauge field $A^a_\mu$ along the inverse tetrad $\tau^\mu$ and $e_a^{\;\,\mu}$ as
\beq
\bal
A_{a\mu} \tau^\mu &=\tau^\mu  \nabla _\mu \left( e_a^{\;\,\nu} a_\nu \right)
-e_a^{\;\,\mu} \nabla _\mu \left( \tau^\nu  a_\nu \right),
\\
A_{a\mu} e_b^{\;\,\mu} &=e_b^{\;\,\mu} A_{ab} -\frac12 \nabla_{[\mu} a_{\nu]} e_a^{\;\,\mu} e_b^{\;\,\nu}.
\eal
\eeq
One can show the equivalence between the actions \eqref{Actioncurvedimp1} and \eqref{S2RAA} by noticing that these relations reduce to \eqref{projA} after the gauge fixing condition \eqref{gfa} is implemented. Similarly, one can define 
\beq
A_{\mu\nu}= A_{a\mu} e^a_{\;\,\nu}-\nabla_\mu a_\nu,
\eeq
which can be shown to be explicitly symmetric after replacing \eqref{solAamucurved}, and matches \eqref{Aamu} after imposing the gauge condition \eqref{gfa}. 

\section{Conclusions and outlook}
\label{sec:conclusions}

In this paper, we have studied the connection between symmetric gauge fields and gravity after understanding MDMA as a contraction of Poincar\'e algebra. This analysis gives a geometric interpretation to the fracton charge as the momentum of the matter field in a transverse (internal) spacetime dimension, and dipole charge as the angular momentum along that direction.

The main result of the paper is twofold: on one hand we have constructed a Lie algebra contraction that allows to obtain the MDMA from the Poincar\'e algebra in one dimension higher by putting together a combination of a pseudo-Carrollian contraction and an Aristotelian one. On the other hand, we have derived the action \eqref{S1S2}, which describes fracton gauge fields coupled to Aristotelian geometry. This action was obtained from a higher-dimensional Poincar\'e gauge theory in a symmetry-broken phase after applying a dimensional reduction and the pseudo-Carrollian-Aristotelian limit.

Different $\sigma\rightarrow 0$ limits of the resulting action were analized which led in particular to the original gauge theory of fracton electrodynamics proposed by Pretko on flat space, together with a flat space Proca extension of the theory, a spontaneously broken phase in curved space, and a symmetric phase in curved space with the harmonic condition $D^\mu  R_{ab\mu\nu} =0$.

As future directions, we envisage possible generalizations of our results. For instance, one could explore the inclusion of fractonic matter in our model by considering higher-dimensional relativistic matter fields coupled to our Poincar\'e gauge theory. Additionally, one could generalize the Lie algebra contraction here considered to fractonic symmetries that generalize the MDMA by including higher moment charges. Indeed, due to the isomorphism between the Bargmann algebra and the extension of the MDMA that includes conservation of the trace of the cuadrupole moment, it would be interesting to understand the relation between Newton-Cartan gravity and fracton gauge theories with such a gauge group. Moreover, it would be of interest to explore supersymmetric generalizations of our results. By exploiting the relation between the Carroll algebra and the MDMA, supersymmetric extensions of Carroll could be of use in the study of spin $1/2$ fractons. 

\acknowledgments
We thank E. Bergshoeff, G. Palumbo, O. Castillo-Felisola for enlightening comments and discussions.  F. P.-B. acknowledges the Nordita Institute for hospitality while attending the workshop "Hydrodynamics at all scales". This work has been funded by the Norwegian Financial Mechanism 2014-2021 via the Narodowe Centrum Nauki (NCN) POLS grant 2020/37/K/ST3/03390.


\bibliographystyle{JHEP}
\bibliography{Draft}

\end{document}